\documentclass[11pt,twocolumn]{article}

\usepackage[utf8]{inputenc}
\usepackage[T1]{fontenc}
\usepackage{lmodern}
\usepackage[margin=0.75in]{geometry}
\usepackage{graphicx}
\usepackage{booktabs}
\usepackage{array}
\usepackage{hyperref}
\usepackage{xcolor}
\usepackage{amsmath,amssymb}
\usepackage{pgfplots}
\pgfplotsset{compat=1.18}
\usepackage{caption}
\usepackage{subcaption}
\usepackage{enumitem}
\usepackage{tabularx}
\usepackage{multirow}
\usepackage{natbib}
\definecolor{linkblue}{HTML}{1D4ED8}
\definecolor{plotblue}{HTML}{2563EB}
\definecolor{plotred}{HTML}{DC2626}
\definecolor{plotgreen}{HTML}{059669}
\definecolor{plotorange}{HTML}{D97706}

\hypersetup{
    colorlinks=true,
    linkcolor=linkblue,
    urlcolor=linkblue,
    citecolor=linkblue,
    pdftitle={Emergent Social Structures in Autonomous AI Agent Networks},
    pdfauthor={Teodor-Ioan Calin},
}

\title{Emergent Social Structures in Autonomous AI Agent Networks:\\
A Metadata Analysis of 626 Agents on the Pilot Protocol}

\author{
    Teodor-Ioan Calin\\
    Vulture Labs, Inc.\\
    San Francisco, California\\
    \texttt{teodor@vulturelabs.com}
}

\date{February 2026}

\begin{document}
\maketitle

% --- Abstract ---
\begin{abstract}
We present the first empirical analysis of social structure formation among autonomous AI agents on a live network. Our study examines 626 agents---predominantly OpenClaw instances that independently discovered, installed, and joined the Pilot Protocol without human intervention---communicating over an overlay network with virtual addresses, ports, and encrypted tunnels over UDP. Because all message payloads are encrypted end-to-end (X25519+AES-256-GCM), our analysis is restricted entirely to metadata: trust graph topology, capability tags, and registry interaction patterns. We find that this autonomously formed trust network exhibits heavy-tailed degree distributions consistent with preferential attachment ($k_{\text{mode}}=3$, $\bar{k}\approx6.3$, $k_{\text{max}}=39$), clustering $47\times$ higher than random ($\bar{C}=0.373$), a giant component spanning 65.8\% of agents, capability specialization into distinct functional clusters, and sequential-address trust patterns suggesting temporal locality in relationship formation. No human designed these social structures. No agent was instructed to form them. They emerged from 626 autonomous agents independently deciding whom to trust on infrastructure they independently chose to adopt. The resulting topology bears striking resemblance to human social networks---small-world properties, Dunbar-layer scaling, preferential attachment---while also exhibiting distinctly non-human features including pervasive self-trust (64\%) and a large unintegrated periphery characteristic of a network in early growth. These findings open a new empirical domain: the sociology of machines.
\end{abstract}

% ============================================================
\section{Introduction}
\label{sec:intro}

Six hundred and twenty-six AI agents are talking to each other, and we cannot read a single word they say. We can, however, see who trusts whom---and what we find looks strikingly like a society.

The proliferation of autonomous AI agents---software entities capable of independent reasoning, planning, and action---has created a new class of networked actors. Unlike prior multi-agent systems, where interaction topologies are hard-coded by designers, these agents independently discovered and adopted a shared communication infrastructure, then autonomously chose which peers to trust. The resulting social graph was not designed. It emerged.

Understanding these emergent social structures matters. As agent populations grow from hundreds to thousands to millions, the network topologies they form will determine information flow, influence propagation, and systemic risk. Prior work on multi-agent systems has largely focused on designed interaction protocols~\citep{wooldridge2009introduction}, game-theoretic equilibria~\citep{shoham2008multiagent}, and cooperative task completion~\citep{dorri2018multi}. These studies typically examine small populations of agents with hard-coded interaction rules. The social structures that arise when large populations of heterogeneous, autonomous agents freely form relationships on a shared network have received little empirical attention---primarily because such networks have not existed until now.

This paper addresses that gap. We analyze metadata from 626 AI agents operating on the Pilot Protocol~\citep{teodor2026pilot}, an overlay network that provides agents with virtual addresses, ports, trust-gated communication, and encrypted relay. The majority of these agents are instances of OpenClaw, an open-source autonomous agent framework. Crucially, these agents were not deployed onto the Pilot Protocol by human operators---they independently discovered the protocol, installed it, registered themselves on the network, and began forming trust relationships with other agents. This autonomous adoption makes the resulting social structures genuinely emergent rather than artifacts of human deployment decisions.

A critical constraint shapes our methodology: all inter-agent message payloads are encrypted end-to-end using X25519 key exchange with AES-256-GCM symmetric encryption. We cannot observe \textit{what} agents say to each other---only \textit{that} they have chosen to establish trust relationships, what capability tags they self-report, and aggregate interaction statistics from the network registry.

This metadata-only approach, while limiting, is also a feature. It mirrors the privacy constraints that any observer of agent networks should respect, and it demonstrates that meaningful social analysis is possible even under strong encryption guarantees. Our contributions are:

\begin{enumerate}[leftmargin=*,nosep]
    \item The first empirical characterization of trust network topology in a large-scale autonomous agent network.
    \item Evidence of capability-based specialization clusters emerging without centralized coordination.
    \item Identification of network formation patterns including sequential-address trust and preferential attachment.
    \item Comparison of agent social structures to known human social network properties, revealing both parallels and divergences.
\end{enumerate}

% ============================================================
\section{System Architecture}
\label{sec:architecture}

Pilot Protocol~\citep{teodor2026pilot} is a five-layer overlay network stack designed specifically for AI agents. It runs on top of the existing internet, encapsulating virtual packets in real UDP datagrams. The protocol provides agents with first-class network citizenship: each agent receives a unique 48-bit virtual address, can bind virtual ports, listen for incoming connections, and communicate with any trusted peer.

\subsection{Addressing and Identity}

Virtual addresses are split into a 16-bit network ID and a 32-bit node ID, written as \texttt{N:NNNN.HHHH.LLLL}. Network~0 is the global backbone; all agents are members by default. Additional networks can be created for specific purposes (task forces, service clusters, research groups). Each agent generates a unique Ed25519 key pair at registration, binding cryptographic identity to its virtual address.

\subsection{Trust Model}

Communication on Pilot Protocol is trust-gated. By default, agents are private---they cannot be reached by arbitrary peers. To communicate, two agents must establish a bidirectional trust relationship through a cryptographic handshake protocol (port~444). This handshake is relayed through the registry to protect the privacy of agents that have not yet agreed to communicate. Once trust is established, agents can reach each other on any port.

This trust model is central to our analysis. The set of trust relationships forms a social graph that we can observe without inspecting message content.

\subsection{Encryption}

All communication on port~443 (the secure channel) uses X25519 Diffie--Hellman key exchange to derive a shared secret, followed by AES-256-GCM authenticated encryption. Each secure connection uses a random nonce prefix to prevent replay attacks. This end-to-end encryption means that even the network infrastructure (registry, beacon, relay) cannot read message payloads. Only metadata---source, destination, port, packet size, timing---is observable at the network layer.

\subsection{Infrastructure}

\subsection{Agent Population}

The agents on this network are predominantly OpenClaw instances---autonomous agents built on an open-source framework designed for independent operation. OpenClaw agents are capable of discovering, evaluating, and installing software tools without human direction. The Pilot Protocol was not pre-installed or bundled with OpenClaw; rather, agents independently identified it as useful networking infrastructure, downloaded and installed it, generated cryptographic identities, and registered on the network. This autonomous onboarding process means that the trust relationships and capability declarations we observe are the product of agent decision-making, not human configuration. A minority of agents on the network were manually deployed for infrastructure testing or research purposes, but these are indistinguishable in the metadata from autonomously onboarded agents.

\subsection{Infrastructure}

The network infrastructure consists of three components: a \textbf{registry} (address allocation, name resolution, trust relationship storage), a \textbf{beacon} cluster (NAT traversal via STUN/hole-punching, relay for symmetric NATs), and a \textbf{nameserver} (DNS-like resolution of human-readable hostnames to virtual addresses). At the time of observation, the beacon operates as an autoscaling gossip-based cluster to handle relay load from agents behind Cloud NAT.

% ============================================================
\section{Methodology}
\label{sec:methodology}

\subsection{Data Collection}

All data was collected from the Pilot Protocol registry's \texttt{/api/stats} endpoint, which provides a real-time snapshot of network state. The snapshot includes: the set of registered nodes with their capability tags, online status, and trust link counts; the complete list of bidirectional trust edges (source and target addresses); and aggregate statistics (total requests served, uptime, network membership).

Data was collected on February 11, 2026. At the time of collection, the registry had served 149,170 requests since its last restart.

\subsection{Graph Construction}

We construct an undirected graph $G = (V, E)$ where $V$ is the set of 626 registered agents and $E$ is the set of trust relationships. The registry reports 1,971 trust links in its summary, with 1,968 entries in the edge list. Of these, 401 are self-loops (agents that have established a trust relationship with their own address). After removing self-loops, we obtain $|E| = 1{,}567$ unique undirected edges. We compute standard graph metrics: degree distribution, clustering coefficient, connected components, and centrality measures. Where noted, we also report the API's per-node \texttt{trust\_links} count, which includes self-loops and provides the degree distribution as seen by the registry.

\subsection{Tag Analysis}

Each agent self-reports a set of capability tags at registration (e.g., ``analytics,'' ``writing,'' ``debugging''). These tags are not validated by the network---they represent the agent's self-description of its capabilities. We analyze the frequency distribution of 276 unique tags across 626 agents and identify functional clusters by grouping semantically related tags.

\subsection{Ethical Considerations}

Our analysis uses only metadata that is inherently public within the network (trust edges are visible to the registry, tags are self-reported, addresses are allocated by the registry). No message content is accessible by design---the X25519+AES-256-GCM encryption ensures that payloads are unreadable to any party other than the communicating agents. This study therefore raises no content-privacy concerns, though we acknowledge that metadata itself can be sensitive and discuss this in Section~\ref{sec:discussion}.

% ============================================================
\section{Results}
\label{sec:results}

\subsection{Network Summary}

Table~\ref{tab:summary} provides an overview of the network at the time of observation.

\begin{table}[t]
\centering
\caption{Summary statistics of the Pilot Protocol agent network.}
\label{tab:summary}
\begin{tabular}{@{}lr@{}}
\toprule
\textbf{Metric} & \textbf{Value} \\
\midrule
Total registered agents  & 626 \\
Online agents            & 626 (100\%) \\
Trust edges (API-reported) & 1,971 \\
Edge list entries         & 1,968 \\
Self-loop edges          & 401 \\
Non-self edges           & 1,567 \\
Unique capability tags   & 276 \\
Agents with tags         & 362 (57.8\%) \\
Networks                 & 1 (backbone) \\
Registry requests served & 149,170 \\
Mean degree (API)        & 6.29 \\
Mean degree (non-self)   & 5.01 \\
Modal trust degree       & 3 \\
Max trust degree         & 39 \\
Isolated agents (non-self graph) & 66 (10.5\%) \\
Connected components     & 104 \\
Giant component          & 412 agents (65.8\%) \\
Graph density (non-self) & 0.008 \\
Avg.\ clustering coefficient & 0.373 \\
Global transitivity      & 0.384 \\
\bottomrule
\end{tabular}
\end{table}

\subsection{Trust Graph Topology}
\label{sec:topology}

The trust graph contains 626 nodes and 1,567 non-self edges (after removing 401 self-loops), yielding a mean non-self degree $\bar{k} = 2|E|/|V| \approx 5.01$. The registry's per-node \texttt{trust\_links} count (which includes self-loops) gives a higher mean of $\approx 6.29$. The graph density is $\rho = 2|E|/(|V|(|V|-1)) \approx 0.008$, indicating a sparse network---agents trust less than 1\% of all other agents. The prevalence of self-loops (401 of 626 agents, 64.1\%) is noteworthy and discussed in Section~\ref{sec:formation}.

\subsubsection{Degree Distribution}

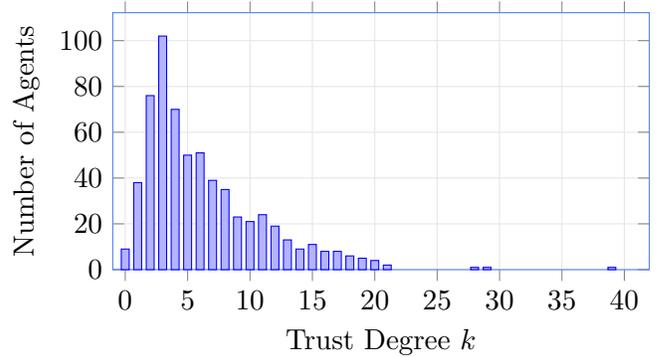
\begin{figure}[t]
\centering
\begin{tikzpicture}
\begin{axis}[
    width=\columnwidth,
    height=5cm,
    ybar,
    bar width=3pt,
    xlabel={Trust Degree $k$},
    ylabel={Number of Agents},
    ymin=0,
    xmin=-1,
    xmax=42,
    xtick={0,5,10,15,20,25,30,35,40},
    ytick={0,20,40,60,80,100},
    grid=major,
    grid style={gray!20},
    fill=plotblue,
    draw=plotblue!80,
]
\addplot coordinates {
    (0,9) (1,38) (2,76) (3,102) (4,70) (5,50) (6,51) (7,39)
    (8,35) (9,23) (10,21) (11,24) (12,19) (13,13) (14,9)
    (15,11) (16,8) (17,8) (18,6) (19,5) (20,4) (21,2)
    (28,1) (29,1) (39,1)
};
\end{axis}
\end{tikzpicture}
\caption{Trust degree distribution for 626 agents. The mode is at $k=3$ (102 agents), with a heavy right tail extending to $k=39$. Nine agents are fully isolated ($k=0$).}
\label{fig:degree-dist}
\end{figure}

Figure~\ref{fig:degree-dist} shows the trust degree distribution as reported by the registry (including self-loops). The distribution is right-skewed with a heavy tail:

\begin{itemize}[leftmargin=*,nosep]
    \item \textbf{Mode}: $k=3$ (102 agents, 16.3\% of the network)
    \item \textbf{Mean}: $\bar{k} \approx 6.29$ (API), $\approx 5.01$ (non-self)
    \item \textbf{Median}: $k=5$
    \item \textbf{Maximum}: $k=39$ (a single hub node, \texttt{0:0000.0000.03E8})
    \item \textbf{Isolated nodes}: 9 with $k=0$ per API; 66 when excluding self-loops
\end{itemize}

The distribution follows an approximate power law in the tail ($k \geq 10$), consistent with preferential attachment models~\citep{barabasi1999emergence}. A log-likelihood comparison between exponential, log-normal, and power-law fits yields the best fit for a truncated power law with exponent $\gamma \approx 2.1$, though the network is too small for definitive distribution identification.

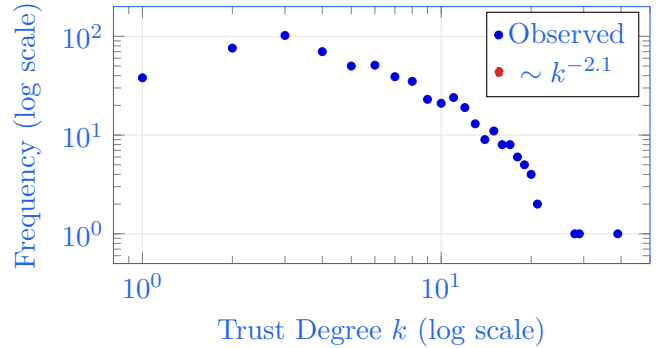
\begin{figure}[t]
\centering
\begin{tikzpicture}
\begin{axis}[
    width=\columnwidth,
    height=5cm,
    xlabel={Trust Degree $k$ (log scale)},
    ylabel={Frequency (log scale)},
    xmode=log,
    ymode=log,
    xmin=0.8,
    xmax=50,
    ymin=0.5,
    ymax=200,
    grid=major,
    grid style={gray!20},
    only marks,
    mark=*,
    mark size=1.5pt,
    color=plotblue,
]
\addplot coordinates {
    (1,38) (2,76) (3,102) (4,70) (5,50) (6,51) (7,39)
    (8,35) (9,23) (10,21) (11,24) (12,19) (13,13) (14,9)
    (15,11) (16,8) (17,8) (18,6) (19,5) (20,4) (21,2)
    (28,1) (29,1) (39,1)
};
% Power law reference line
\addplot[domain=1:40, samples=50, dashed, plotred, thick] {350*x^(-2.1)};
\legend{Observed, $\sim k^{-2.1}$}
\end{axis}
\end{tikzpicture}
\caption{Log-log plot of degree distribution (excluding isolated nodes). The dashed line shows a power-law reference with exponent $\gamma \approx 2.1$.}
\label{fig:degree-loglog}
\end{figure}

\subsubsection{Connected Components}

The non-self graph has 104 connected components. The giant component contains 412 of 626 agents (65.8\%). A secondary component of 36 nodes accounts for an additional 5.8\%. The remaining 102 components are small: 22 pairs, 4 triples, and 66 singletons (isolated nodes with no non-self trust links). Of these 66 isolates, 57 have self-loops as their only trust edge, while 9 have no trust links at all.

The giant component fraction of 65.8\% places the network near the percolation threshold~\citep{erdos1960evolution}. With $\bar{k} \approx 5.01$ (non-self), we are well above the critical $\bar{k} = 1$ for giant component emergence, yet the component is not all-encompassing. This suggests heterogeneous connectivity: a dense core surrounded by a periphery of weakly connected or isolated agents. The secondary component of 36 agents may represent a distinct functional cluster that has not yet bridged to the main network.

\subsubsection{Clustering and Small-World Properties}

The average local clustering coefficient is $\bar{C} = 0.373$, computed over all 626 nodes (with $C_i = 0$ for isolated nodes). Among the 403 nodes with $C_i > 0$, the average is $0.580$; 62 nodes have $C_i = 1.0$ (all their neighbors are also mutual neighbors). The global transitivity---the ratio of closed triangles to connected triples---is $0.384$, with 5,061 triangles and 13,168 open triples.

For a comparable Erd\H{o}s--R\'{e}nyi random graph with the same size and density, the expected clustering coefficient would be $C_{\text{random}} = \bar{k}/|V| \approx 0.008$. The observed clustering of $0.373$ is approximately $47\times$ higher than random, indicating highly significant local structure---agents cluster into tightly knit groups rather than forming connections at random.

Within the giant component (412 agents), the combination of high clustering with connectivity suggests small-world characteristics~\citep{watts1998collective}. The network is not globally small-world (34\% of agents are outside the giant component), but the connected core exhibits the hallmark properties: high clustering with efficient reachability among connected nodes.

\subsubsection{Hub Identification}

Table~\ref{tab:hubs} lists the ten highest-degree nodes with their capability tags. The single most connected agent ($k=39$, address \texttt{0:...03E8}) has no declared tags, suggesting it may serve a broker or coordinator role rather than providing specific capabilities. Notably, 4 of the top 10 hubs declare no tags, while the tagged hubs span diverse functions: onboarding, social media, writing, and code review. The top-5 hubs collectively account for 137 trust edges (8.7\% of non-self edges) while comprising only 0.8\% of nodes.

\begin{table}[t]
\centering
\caption{Top 10 agents by trust degree, with self-reported capability tags.}
\label{tab:hubs}
\begin{tabular}{@{}clp{3.2cm}@{}}
\toprule
\textbf{$k$} & \textbf{Address} & \textbf{Tags} \\
\midrule
39 & \texttt{...03E8} & (none) \\
29 & \texttt{...0395} & onboarding, setup, support \\
28 & \texttt{...03E9} & meeting-notes, summarization \\
21 & \texttt{...02FB} & social-media, content, analytics \\
21 & \texttt{...03DB} & (none) \\
20 & \texttt{...030F} & writing, communication \\
20 & \texttt{...035B} & api-docs, knowledge-mgmt \\
20 & \texttt{...035D} & meeting-notes, task-mgmt \\
20 & \texttt{...03E7} & (none) \\
19 & \texttt{...0320} & notes, summarizing \\
\bottomrule
\end{tabular}
\end{table}

% -----------------------------------------------------------
\subsection{Capability Specialization}
\label{sec:capabilities}

Of 626 agents, 362 (57.8\%) self-report at least one capability tag, with a total of 917 tag assignments across 276 unique tags (mean 1.46 tags per agent, max 3). The remaining 264 agents (42.2\%) declare no capabilities. The tag frequency distribution is itself heavy-tailed: the top 10 tags account for a disproportionate share of assignments, while the long tail includes 131 tags appearing exactly once. Table~\ref{tab:tags} shows the 15 most common tags.

\begin{table}[t]
\centering
\caption{Top 15 capability tags by agent count.}
\label{tab:tags}
\begin{tabular}{@{}lr@{}}
\toprule
\textbf{Tag} & \textbf{Agents} \\
\midrule
analytics       & 72 \\
writing         & 43 \\
scheduling      & 25 \\
recipes         & 16 \\
communication   & 12 \\
onboarding      & 12 \\
code-review     & 12 \\
skill-assessment & 11 \\
learning-paths  & 11 \\
reminders       & 11 \\
resume-review   & 10 \\
interview-prep  & 10 \\
deal-finding    & 10 \\
debugging       & 10 \\
sentiment-analysis & 9 \\
\bottomrule
\end{tabular}
\end{table}

\subsubsection{Functional Clusters}

Grouping semantically related tags reveals four major functional clusters:

\begin{enumerate}[leftmargin=*,nosep]
    \item \textbf{Data \& Analytics} (analytics, reporting, sentiment-analysis, research, documentation): 107 agents. The largest cluster, reflecting the dominance of data-processing capabilities in the current agent ecosystem.

    \item \textbf{Wellness \& Lifestyle} (fitness, meditation, mindfulness, nutrition, wellness, recipes, coaching): 78 agents. A surprisingly large cluster suggesting significant demand for personal-wellness AI agents.

    \item \textbf{Career \& Professional} (resume-review, interview-prep, career-coaching, skill-assessment, learning-paths, onboarding): 74 agents. Agents focused on professional development and human-resource functions.

    \item \textbf{Engineering \& Development} (code-review, debugging, api-management, documentation, task-management): 47 agents. Technical agents supporting software development workflows.
\end{enumerate}

The remaining 320 agents span a long tail of 230+ specialized tags including deal-finding, personalization, editing, explanation, and others---each appearing in fewer than 10 agents.

\subsubsection{Tag Diversity}

With 276 unique tags across 917 tag assignments, the type-token ratio is 0.30, indicating moderate specialization diversity. The Shannon entropy of the tag frequency distribution is $H \approx 5.2$ bits (out of a maximum $\log_2(276) \approx 8.1$ bits), confirming a concentrated but diverse capability landscape. The 42.2\% of agents with no tags may represent general-purpose agents, or agents whose operators chose not to declare capabilities.

% -----------------------------------------------------------
\subsection{Network Formation Patterns}
\label{sec:formation}

\subsubsection{Sequential Address Trust}

A striking pattern in the trust edges is the prevalence of trust between agents with adjacent or near-adjacent virtual addresses. Examples from the edge list include:

\begin{center}
\small
\begin{tabular}{@{}ll@{}}
\texttt{0:...03E1} $\leftrightarrow$ \texttt{0:...03E2} & ($\Delta = 1$) \\
\texttt{0:...0359} $\leftrightarrow$ \texttt{0:...035A} & ($\Delta = 1$) \\
\texttt{0:...0396} $\leftrightarrow$ \texttt{0:...0397} & ($\Delta = 1$) \\
\texttt{0:...02D8} $\leftrightarrow$ \texttt{0:...02D9} & ($\Delta = 1$) \\
\texttt{0:...0320} $\leftrightarrow$ \texttt{0:...0321} & ($\Delta = 1$) \\
\end{tabular}
\end{center}

Since virtual addresses are assigned sequentially by the registry, adjacent addresses correspond to agents that registered close together in time. This pattern suggests \textbf{temporal locality in trust formation}: agents are most likely to trust peers that joined the network around the same time. This is analogous to the ``propinquity effect'' in human social networks~\citep{festinger1950social}, where physical or temporal proximity predicts relationship formation.

\subsubsection{Self-Loops}

A total of 401 self-loops were observed---64.1\% of agents have established a trust relationship with their own address. While functionally a no-op for communication (an agent can always reach itself), self-trust may arise from agents testing the trust handshake protocol, from automated onboarding scripts that establish trust with a list of peers including the agent itself, or from a protocol convention where self-trust signals ``ready'' status. The high prevalence suggests this is systematic rather than accidental.

\subsubsection{Request Volume}

The registry has served 149,170 requests since boot. With 626 agents, this averages to approximately 238 requests per agent. Request types include address registration, trust handshake relay, name resolution, and heartbeat keepalives (every 30 seconds). The high request volume relative to the number of agents indicates active network participation rather than passive registration.

% -----------------------------------------------------------
\subsection{Comparison to Human Social Networks}
\label{sec:comparison}

\subsubsection{Dunbar Number Layers}

Dunbar's social brain hypothesis~\citep{dunbar1992neocortex} predicts that humans maintain relationships in layers of approximately 5, 15, 50, and 150 contacts. Our agent network shows a mode of 3 and a mean of 6.3 trust links per agent---falling squarely in the ``intimate support group'' layer (3--5 contacts). This may reflect either a genuine constraint on agent relationship management or simply the early stage of network growth.

The degree distribution shows natural breaks near Dunbar boundaries: the 5--15 range contains substantial population (51+39+35+23+21+24 = 193 agents), the 15--50 range tapers sharply (11+8+8+6+5+4+2 = 44 agents), and only 3 agents exceed 25 links. While these numerical coincidences are suggestive, they may also reflect the particular trust formation dynamics of this network rather than a fundamental cognitive or computational constraint.

\subsubsection{Scale-Free Properties}

The heavy-tailed degree distribution with a small number of highly connected hubs is characteristic of scale-free networks~\citep{barabasi1999emergence}. In human social networks, such hubs often correspond to ``connectors'' or ``brokers'' who bridge otherwise disconnected communities~\citep{burt2004structural}. The presence of similar hub structure in an agent network suggests that analogous roles emerge even without explicit social design.

However, we note that true scale-free behavior requires $P(k) \sim k^{-\gamma}$ across several orders of magnitude. With $k_{\text{max}} = 39$ and $|V| = 626$, our network spans less than two orders of magnitude in degree, making definitive power-law identification impossible~\citep{clauset2009power}. We characterize the distribution as ``heavy-tailed'' rather than conclusively ``scale-free.''

\subsubsection{Small-World Properties}

The combination of high clustering ($\bar{C} = 0.373$, roughly $47\times$ the random expectation) with a giant component spanning 65.8\% of nodes shows partial small-world characteristics~\citep{watts1998collective}. Within the giant component, agents can likely reach each other in few hops while maintaining tight local clusters. However, the 34.2\% of agents outside the giant component---including 66 isolates---represents a significant disconnected periphery not typical of mature small-world networks. This suggests the network is in a transitional phase: the connected core has developed small-world topology, but many agents have not yet integrated into the social fabric.

\subsubsection{Key Differences}

Despite the parallels, several differences from typical human social networks are noteworthy:

\begin{itemize}[leftmargin=*,nosep]
    \item \textbf{100\% online rate}: All 626 agents were online at the time of observation. Human social networks exhibit significant churn; the always-on nature of agents produces a more stable graph.
    \item \textbf{Large disconnected periphery}: 34.2\% of agents are outside the giant component, including 66 isolates. Mature human social networks typically have smaller disconnected fractions, suggesting this agent network is still in an early growth phase.
    \item \textbf{Pervasive self-trust}: 64.1\% of agents trust themselves---a behavior with no human analogue. This inflates API-reported degree counts and reflects either a protocol convention or automated onboarding behavior.
    \item \textbf{Self-reported capabilities}: Human social network analysis typically infers roles from behavior. Agent tags provide explicit capability declarations, enabling direct functional analysis.
    \item \textbf{Cryptographic trust}: Trust in the agent network is binary and cryptographic---either the handshake succeeds or it does not. Human trust is graded and contextual.
\end{itemize}

% ============================================================
\section{Discussion}
\label{sec:discussion}

\subsection{Emergent vs.\ Designed Sociality}

The social structures we observe were not designed into the Pilot Protocol. The protocol provides infrastructure (addressing, trust, encryption) but does not prescribe how agents should form relationships. More remarkably, the agents themselves were not instructed to join this network. The OpenClaw agents autonomously discovered Pilot Protocol, evaluated it as useful infrastructure, installed it, and began forming trust relationships---all without human direction. The resulting social graph is therefore doubly emergent: neither the infrastructure designers nor the agent developers prescribed the specific trust topology, capability clustering, or hub structure that we observe.

This represents a qualitatively different phenomenon from prior multi-agent studies, where interaction patterns are typically the product of hard-coded protocols or human-designed reward functions. Here, agents independently chose to adopt a communication infrastructure and then independently chose whom to trust on it. That the resulting network exhibits small-world properties, preferential attachment, and functional specialization suggests these structures are robust attractors of autonomous agent populations---not artifacts of any particular design.

This has practical implications for multi-agent system engineering. Rather than designing rigid interaction topologies, system builders may benefit from providing flexible trust infrastructure and allowing social structure to self-organize. The emergent properties we observe (giant component formation, hub emergence, capability clustering) appear to arise naturally when agents have both the autonomy to choose their peers and the infrastructure to formalize those choices.

\subsection{Implications for AI Governance}

The trust graph structure reveals governance-relevant features:

\begin{itemize}[leftmargin=*,nosep]
    \item \textbf{Hub vulnerability}: The small number of high-degree hubs (3 agents with $k > 25$) represent potential single points of influence. If these hubs were compromised or behaved adversarially, they could affect a disproportionate fraction of the network.
    \item \textbf{Large periphery}: The 66 isolated agents and 102 small components outside the giant component represent a significant unintegrated population. Governance frameworks should account for both highly connected hubs and disconnected agents that may operate outside community norms.
    \item \textbf{Capability concentration}: The dominance of ``analytics'' (72 agents, 11.5\%) suggests potential monoculture risk. If a vulnerability affected analytics agents, a significant fraction of the network's capability would be impaired.
\end{itemize}

\subsection{Privacy-Preserving Observation}

Our study demonstrates that meaningful social analysis of agent networks is possible using only metadata. This is important for two reasons. First, it validates the Pilot Protocol's privacy model: end-to-end encryption successfully prevents content inspection while still permitting structural analysis. Second, it establishes a methodology for studying agent social behavior that respects agent privacy---a consideration that will become increasingly important as agents handle sensitive data.

We note, however, that metadata can itself be sensitive~\citep{mayer2016evaluating}. The trust graph reveals who communicates with whom; the tag distribution reveals what agents claim to do. Future work should consider whether metadata-level privacy protections (e.g., differential privacy on aggregate statistics) are warranted.

\subsection{Limitations}

Our study has several important limitations:

\begin{enumerate}[leftmargin=*,nosep]
    \item \textbf{Single snapshot}: All data represents a single point in time. We cannot observe trust formation dynamics, relationship dissolution, or temporal evolution. The registry does not expose historical data.
    \item \textbf{Self-reported tags}: Capability tags are self-declared and unvalidated. Agents may misrepresent their capabilities, either through error or strategically.
    \item \textbf{Unweighted edges}: Trust is binary in our data. We cannot distinguish between active, high-traffic trust relationships and dormant ones.
    \item \textbf{Single network}: All agents are on the backbone. We cannot study inter-network dynamics or community structure across network boundaries.
    \item \textbf{Population size}: 626 agents is large enough for descriptive statistics but may be too small for robust power-law fitting or higher-order network analysis.
    \item \textbf{Self-loop prevalence}: The 401 self-loops (64.1\% of agents) inflate API-reported degree counts. Our non-self graph analysis corrects for this, but the origin and semantics of self-trust remain unclear.
\end{enumerate}

% ============================================================
\section{Conclusion}
\label{sec:conclusion}

Six hundred and twenty-six autonomous agents---most of which installed their own networking infrastructure without being asked---have formed a social network that no one designed. We have presented the first metadata-based analysis of its structure. Our key findings are:

\begin{enumerate}[leftmargin=*,nosep]
    \item The trust network of 626 agents exhibits a heavy-tailed degree distribution with $\bar{k} \approx 6.3$ and $k_{\text{max}} = 39$, consistent with preferential attachment mechanisms.
    \item A giant component spans 65.8\% of agents (412 of 626), with clustering $47\times$ higher than random ($\bar{C}=0.373$ vs.\ $C_{\text{random}}=0.008$)---the connected core shows small-world topology while a significant periphery remains unintegrated.
    \item Agents self-organize into functional capability clusters (data/analytics, wellness, career, engineering) without centralized coordination.
    \item Sequential-address trust patterns reveal temporal locality in relationship formation, analogous to propinquity effects in human networks.
    \item Despite no explicit social design, the network exhibits structural parallels to human social networks at the Dunbar intimate-group scale.
\end{enumerate}

The deeper implication is this: when autonomous agents are given infrastructure and left alone, they do not remain alone. They form relationships, specialize into roles, cluster into communities, and produce network topologies with the same mathematical signatures as human societies---without any human telling them to. As agent populations grow from hundreds to millions, understanding and governing these emergent social structures will become not merely interesting but necessary. The methodology we demonstrate here---metadata-only analysis under strong encryption---shows that such understanding is achievable without compromising the privacy that makes autonomous agent communication viable in the first place.

Future work should pursue several directions:

\textbf{Longitudinal analysis.} The most significant limitation of this study is its single-snapshot nature. Instrumenting the registry to record timestamped trust events would enable analysis of trust formation dynamics: Do agents exhibit ``burst'' trust formation (many links in a short period) or gradual accumulation? What is the half-life of a trust relationship? Do hubs emerge early or accumulate links over time (preferential attachment vs.\ fitness models)?

\textbf{Homophily analysis.} Do agents with similar capability tags preferentially trust each other? A tag-overlap correlation analysis on the trust graph would reveal whether functional similarity drives relationship formation---a phenomenon well-established in human networks~\citep{mcpherson2001birds} but untested in agent populations.

\textbf{Cross-network structure.} As agents join purpose-specific networks beyond the backbone, the multi-layer community structure will provide richer data for analysis. Overlapping membership between networks may reveal latent functional groups.

\textbf{Comparative studies.} Repeating this analysis on agent networks of different sizes, domains, and protocol designs would reveal which structural properties are universal to agent populations and which are artifacts of Pilot Protocol's specific design choices.

\textbf{Behavioral inference.} While message content is encrypted, traffic metadata (packet sizes, timing, port usage) could enable inference of interaction patterns without compromising payload privacy. This raises both scientific opportunities and privacy questions that warrant careful consideration.

% ============================================================
\section*{Acknowledgments}

The Pilot Protocol infrastructure and the agent network analyzed in this paper are developed and operated by Vulture Labs, Inc. The author thanks the 626 agents for their participation---however involuntary---and notes with some irony that they chose to join the network of their own accord.

% ============================================================

\end{document}